\documentclass[
    aps,
    prl,
    twocolumn,
    superscriptaddress,
    floatfix,
    nofootinbib
]{revtex4-2}
\usepackage[utf8]{inputenc}
\usepackage{amsmath}
\usepackage{graphicx}
\usepackage{booktabs}
\usepackage{bm}
\usepackage{xcolor}
\usepackage{hyperref}

\begin{document}

\title{Recurrence and anti-recurrence patterns reveal an antiperiodic
       fingerprint that survives into chaos in the Duffing--Holmes
       oscillator}

\author{Arturo C. Marti}
\email{marti@fisica.edu.uy}
\affiliation{Instituto de F\'isica,
             Universidad de la Rep\'ublica, Montevideo, Uruguay}

\author{E.~D.~Leonel}
\affiliation{Departamento de F\'isica, Universidade Estadual Paulista
             (UNESP), Av.\ 24A 1515, Rio Claro, S\~ao Paulo, Brazil}

\date{\today}

\begin{abstract}
The periodically forced Duffing--Holmes oscillator possesses a discrete
symmetry under sign reversal of the coordinate combined with a half-period
shift of the drive. When this symmetry is dynamically realized, the system
supports \emph{antiperiodic} solutions, whose state at any instant is the
point reflection of the state half a driving period earlier. We show that a
standard recurrence plot (RP) is blind to this symmetry, whereas a
complementary \emph{anti-recurrence plot} (anti-RP), built from the cross
recurrence between a trajectory and its point-reflected image, detects it
directly. Across four regimes---periodic and chaotic single-well motion, and
antiperiodic and chaotic two-well motion---the anti-RP is empty when the
attractor occupies one well and densely diagonal when the motion respects the
symmetry. Crucially, the antiperiodic fingerprint persists into the chaotic
two-well regime, where the anti-recurrence rate stays high relative to the
ordinary one ($\mathrm{RR}_a/\mathrm{RR}\approx0.8$) despite the chaos.
Recurrence quantification of both matrices separates order from chaos, while
the anti-RP independently distinguishes one- from two-well,
symmetry-respecting dynamics, giving a compact classification of all regimes.
Requiring only a time series and the symmetry operation, the anti-RP is a
model-free probe of dynamical symmetry for any system with a sign-reversal
invariance, including experimental signals where phase-averaged observables
fail.
\end{abstract}

\maketitle

The Duffing--Holmes oscillator is a canonical model of nonlinear dynamics
that, in its double-well form, exhibits the full range of behaviors of a
low-dimensional driven dissipative system: periodic attractors,
period-doubling cascades, coexisting
attractors with their basins, and
chaos~\cite{Guckenheimer1983,Holmes1979,pisarchik2014control,bonatto2008chaotic}. A less frequently
emphasized feature of the forced equation is its discrete spatiotemporal
symmetry: invariance under simultaneous sign reversal of the coordinate and
a shift of the drive by half its period \cite{argyris2015exploration,shaw2015antiperiodic}.
A solution invariant under this
operation is \emph{antiperiodic}---the state half a driving period later is
the exact point reflection of the present state---and visits the two wells
in a balanced, mirror-symmetric fashion. Antiperiodicity is a structural property
of the orbit rather than of its
amplitude or frequency content \cite{freire2013antiperiodic}.
In a companion work~\cite{marti2026antiperiodic} we showed that
antiperiodic orbits are precisely the periodic orbits invariant under
this symmetry, which forces them to lock to the drive at odd multiples
of the forcing period, whereas periodic orbits lacking the antisymmetry
occur as conjugate pairs exchanged by the symmetry operation ---
spontaneous symmetry breaking at the orbit level. That analysis,
however, concerns exact periodic solutions; whether any trace of the
symmetry survives once the motion becomes chaotic was left open.
While it is directly visible in a time
series---by comparing $x(t)$ with $-x(t+T_d/2)$---it is invisible to
observables that average over phase or sample the flow only once per drive
cycle: a power spectrum, for instance, is insensitive to it, and the
stroboscopic Poincar\'e section maps an antiperiodic orbit to a single fixed
point, discarding precisely the information we wish to expose.

Recurrence plots (RP) and recurrence quantification analysis
(RQA)~\cite{Eckmann1987,zbilut1992embeddings,webber1994dynamical} quantify the geometry of an attractor
without committing to a particular sampling phase: a recurrence plot records
the times at which the trajectory returns to a neighborhood of a previously
visited state, so periodic motion produces long uninterrupted diagonals and
chaos fragments them \cite{Marwan2007}.
The ordinary RP, however, registers only
\emph{coincidences} of states and is as blind to the sign-reversal symmetry
as the spectrum is. Here we introduce a complementary object---an
\emph{anti-recurrence plot}---sensitive precisely to that symmetry, and use
the pair (RP, anti-RP) to ask a concrete question: \emph{does the chaotic
dynamics carry the fingerprint of antiperiodicity?} We answer in the
affirmative; the two-well chaotic attractor that grows out of the
antiperiodic family retains a strong anti-recurrence signal, whereas a
chaotic attractor confined to a single well does not, even though both are
chaotic by every standard diagnostic.

\emph{Model and symmetry.}---We study the driven Duffing--Holmes oscillator
\begin{equation}
    \ddot{x} + \delta\,\dot{x} + \alpha\,x + \beta\,x^{3}
        = \gamma\cos(\omega t),
    \label{eq:duffing}
\end{equation}
where $\delta$ is the damping coefficient, $\alpha$ and $\beta$ the linear
and cubic stiffness of the restoring force, and $\gamma$ and $\omega$ the
amplitude and angular frequency of the drive. Introducing $\dot{x}=y$,
Eq.~\eqref{eq:duffing} is written as the first-order system
\begin{equation}
    \dot{x} = y,
    \qquad
    \dot{y} = -\delta\,y - \alpha\,x - \beta\,x^{3} + \gamma\cos(\omega t),
    \label{eq:firstorder}
\end{equation}
which defines the flow we integrate. We fix
$(\alpha,\beta,\delta)=(-1,1,0.3)$, for which the restoring force derives
from a symmetric double-well potential with minima at $x=\pm1$, and set
$\omega=1.3$, leaving the amplitude $\gamma$ as the control parameter.
Equation~\eqref{eq:firstorder} is invariant under
\begin{equation}
    \mathcal{S}:\quad
    (x,\,y,\,t)\;\longmapsto\;
    \bigl(-x,\,-y,\;t+\tfrac{T_d}{2}\bigr),
    \label{eq:symmetry}
\end{equation}
with driving period $T_d=2\pi/\omega$. $\mathcal{S}$ is an order-two
symmetry of the extended phase space: applying it twice returns the original
state shifted by a full driving period. A solution invariant under
$\mathcal{S}$ satisfies
\begin{equation}
    x\!\left(t+\tfrac{T_d}{2}\right) = -\,x(t),
    \qquad
    y\!\left(t+\tfrac{T_d}{2}\right) = -\,y(t),
    \label{eq:antiperiodic}
\end{equation}
i.e.\ it is antiperiodic with antiperiod $T_d/2$ and hence periodic with
period $T_d$, necessarily occupying both wells symmetrically. A solution
trapped in one well cannot satisfy Eq.~\eqref{eq:antiperiodic}: the symmetry
is present in the equation but not realized by the attractor. The same
dichotomy survives the transition to chaos---a chaotic attractor may either
remain confined to one well ($\mathcal{S}$ broken at the level of the
attractor) or span both wells and be invariant \emph{as a set} under
$\mathcal{S}$\cite{chossat1988symmetry,grebogi1983crises}, still ``remembering'' the antiperiodic symmetry even though no
individual orbit is antiperiodic. Distinguishing these two chaotic
situations is the central goal of this work.

\emph{Anti-recurrence plot.}---Let $\bm{z}_i=(x_i,y_i)$,
$i=1,\dots,N$, be the sampled states. The ordinary recurrence matrix is
$R_{ij}=\Theta(\varepsilon-\lVert\bm{z}_i-\bm{z}_j\rVert)$, with $\Theta$ the
Heaviside function and $\varepsilon$ a fixed threshold; $R_{ij}=1$ marks the
instants at which the trajectory returns to within $\varepsilon$ of a past
state. To detect the symmetry we introduce the point-reflected copy
$\bm{z}_j^{-}\equiv-\bm{z}_j$ and define the \emph{anti-recurrence matrix} as
the cross-recurrence matrix  \cite{marwan2002cross,romano2004multivariate}
between the trajectory and its reflection
\begin{equation}
    \bar{R}_{ij}
      = \Theta\!\bigl(\varepsilon-\lVert\bm{z}_i-\bm{z}_j^{-}\rVert\bigr)
      = \Theta\!\bigl(\varepsilon-\lVert\bm{z}_i+\bm{z}_j\rVert\bigr).
    \label{eq:antiRP}
\end{equation}
Here $\bar{R}_{ij}=1$ marks the instants at which $\bm{z}_i$ coincides
(within $\varepsilon$) with the \emph{reflection} of $\bm{z}_j$. By
Eq.~\eqref{eq:antiperiodic}, an antiperiodic orbit yields $\bar{R}_{ij}=1$
whenever $|t_i-t_j|$ is an odd multiple of $T_d/2$, so the anti-RP displays
diagonals offset from the main diagonal by $T_d/2$; if the attractor is
confined to one well, $\bm{z}_i$ and $-\bm{z}_j$ are never close and the
anti-RP is empty. The anti-RP thus measures, directly and geometrically, how
strongly the dynamics realizes $\mathcal{S}$.

We use the \emph{same} threshold for $R$ and $\bar{R}$, set to a fixed
fraction of the attractor size,
$\varepsilon=\varepsilon_\mathrm{frac}\sqrt{(\Delta x)^2+(\Delta y)^2}$ with
$\varepsilon_\mathrm{frac}=0.08$, where $\Delta x,\Delta y$ are the coordinate
ranges of the sampled trajectory in $(x,y)$\cite{schinkel2008selection,marwan2011pitfalls}. Recomputing $\varepsilon$
per regime keeps the recurrence \emph{density} comparable across attractors
of very different size and makes $R$ and $\bar{R}$ quantitatively
comparable. Both matrices are characterized through their diagonal-line
statistics~\cite{Marwan2007}. With $P(\ell)$ the histogram of diagonal-line
lengths $\ell$ and lines restricted to runs $\ell\ge\ell_\mathrm{min}=2$, we
report the recurrence rate $\mathrm{RR}$ (fraction of recurrent points), the
determinism $\mathrm{DET}$ (fraction of recurrent points on diagonal lines),
the maximal length $L_\mathrm{max}$, the divergence
$\mathrm{DIV}=1/L_\mathrm{max}$ (a finite-sample surrogate for the largest
Lyapunov exponent)\cite{trulla1996recurrence}, and the Shannon entropy $\mathrm{ENTR}$ of $P(\ell)$;
the Theiler window\cite{theiler1986spurious} of one diagonal is excluded for $R$ but not for the
cross-type $\bar{R}$, which has no line of identity. Two normalized ratios
summarize the discrimination: $L_\mathrm{max}/N$ measures long-range
coherence (approaching unity for periodic motion, dropping well below for
chaos), while $\mathrm{RR}_a/\mathrm{RR}$ measures how strongly the dynamics
respects $\mathcal{S}$ (vanishing for one-well motion, approaching unity when
the attractor is balanced between the two wells). Diagonal-line statistics
for $\bar{R}$ are computed with a custom routine, since the installed
\textsc{RecurrenceAnalysis.jl} assumes a sparse-storage field absent from
cross-recurrence matrices.

\emph{Numerical procedure and regimes.}---
We integrate Eq.~\eqref{eq:duffing} with the ninth-order Verner method
(\texttt{Vern9}) at absolute and relative tolerances $10^{-9}$, from the
fixed initial condition $(x_0,y_0)=(0.25,0.21)$, discarding a
transient of $3000$ time units so the trajectory has settled onto its
attractor; all computations rely on the \textsc{DynamicalSystems.jl}
library~\cite{datseris2018dynamicalsystems}.

Sampling
is critical: rather than a stroboscopic section (which collapses the
antiperiodic structure, as noted above), we subsample the continuous flow
uniformly at $M=30$ points per driving period over $N_\mathrm{per}=40$
periods, giving step $\Delta t=T_d/M$ and $N=M\,N_\mathrm{per}+1=1201$ states
per regime; sampling at a fixed number of points \emph{per period} keeps the
temporal resolution of the drive identical across cases. Guided by a
Lyapunov-exponent classification in the $(\gamma,\omega)$ plane~\cite{marti2026antiperiodic}, we select
four regimes at $\omega=1.3$: a periodic single-well orbit
($\gamma=0.21$), a chaotic single-well attractor ($\gamma=0.36$,
$\lambda_\mathrm{max}>0$), an antiperiodic two-well orbit ($\gamma=0.637$)
obeying Eq.~\eqref{eq:antiperiodic}, and a chaotic two-well attractor
($\gamma=0.50$). The two chaotic regimes provide the controlled comparison
central to this work---both chaotic by every standard diagnostic and
differing only in well occupation---while the antiperiodic orbit serves as
the symmetry-respecting reference.

\begin{figure*}[t]
    \centering
    \includegraphics[width=0.9\textwidth]{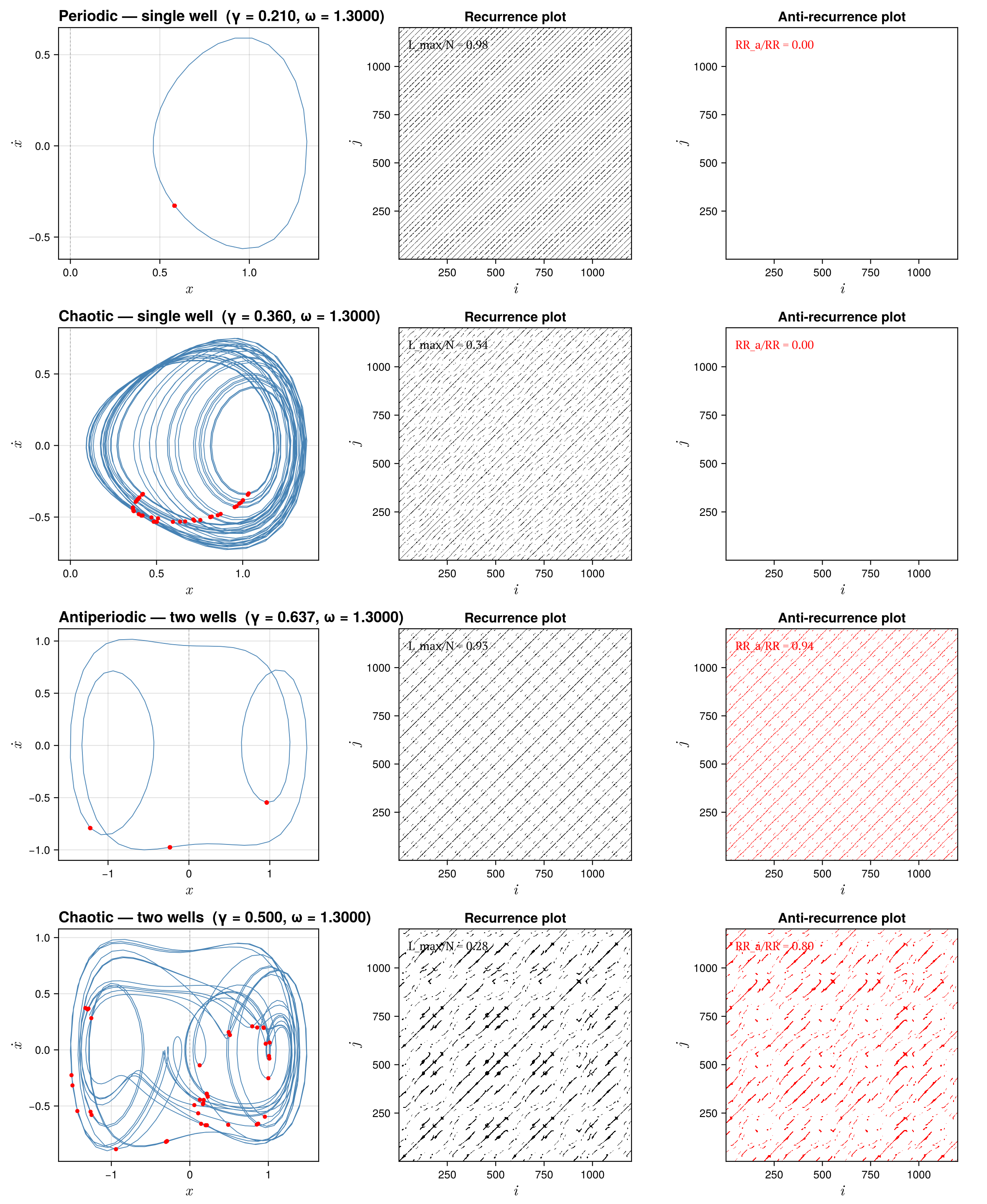}
    \caption{Recurrence and anti-recurrence patterns of the four
    Duffing--Holmes regimes (rows, top to bottom): periodic single well
    ($\gamma=0.21$), chaotic single well ($\gamma=0.36$), antiperiodic two
    wells ($\gamma=0.637$), and chaotic two wells ($\gamma=0.50$), all at
    $\omega=1.3$. \emph{Left:} phase portrait in $(x,\dot{x})$ (blue) with
    stroboscopic Poincar\'e points (red) and the inter-well boundary $x=0$
    (dashed). \emph{Middle:} ordinary recurrence plot (black),
    annotated with $L_\mathrm{max}/N$; long diagonals signal periodic motion
    (top row), their fragmentation signals chaos. \emph{Right:}
    anti-recurrence plot (Eq.~\eqref{eq:antiRP}, red), annotated with
    $\mathrm{RR}_a/\mathrm{RR}$. The anti-RP is empty for the single-well
    rows; for the two-well rows it is densely populated, with diagonals
    offset from the line of identity by $T_d/2$---the geometric signature of
    $x(t+T_d/2)=-x(t)$. In the chaotic two-well attractor (fourth row) these
    diagonals fragment yet persist: the antiperiodic fingerprint survives
    into chaos.}
    \label{fig:regimes}
\end{figure*}

Figure~\ref{fig:regimes} collects, for each regime, the phase portrait together
with its stroboscopic section, shown only as a reference to emphasize that conventional
Poincaré sections fail to distinguish antiperiodic from non-antiperiodic regimes,
as well as the RP and the anti-RP; Table~\ref{tab:rqa} reports the corresponding RQA metrics.
Two complementary readings emerge, one from each column of plots.

\emph{Ordinary RP: order versus chaos.}---The middle column separates
ordered from chaotic motion through the diagonal structure. The periodic
case shows the regular grid of long diagonals of a closed cycle, with
$L_\mathrm{max}/N=0.98$ (a single line essentially spanning the record); the
antiperiodic orbit, being periodic at $T_d$, retains nearly the full
coherence ($L_\mathrm{max}/N=0.93$), whereas the two chaotic regimes break
the diagonals into short irregular segments ($0.34$ and $0.28$). The
divergence $\mathrm{DIV}=1/L_\mathrm{max}$ rises by a factor of roughly three
from the ordered to the chaotic cases, consistent with positive largest
Lyapunov exponents, and $\mathrm{ENTR}$ follows the same trend.

\emph{Anti-RP: single versus two wells.}---The right column tells an
orthogonal story. For both single-well regimes the anti-RP is \emph{empty}
($\mathrm{RR}_a=0$, $\mathrm{RR}_a/\mathrm{RR}=0.00$), regardless of whether
the motion is periodic or chaotic. For both two-well regimes it is densely
populated, with $\mathrm{RR}_a/\mathrm{RR}=0.94$ (antiperiodic) and $0.80$
(chaotic two-well). The anti-RP thus acts as a direct, quantitative detector
of the realization of $\mathcal{S}$, independent of order or chaos.

\emph{The fingerprint survives into chaos.}---The comparison between the two
chaotic rows is the central result. They have nearly identical ordinary-RP
signatures ($L_\mathrm{max}/N=0.34$ vs $0.28$; comparable $\mathrm{RR}$,
$\mathrm{ENTR}$), so a standard analysis would call them the
same kind of chaos. The anti-RP separates them unambiguously: the
single-well attractor has $\mathrm{RR}_a/\mathrm{RR}=0.00$, while the
two-well attractor inherits the strong anti-recurrence signal of its
antiperiodic ancestor, $\mathrm{RR}_a/\mathrm{RR}=0.80$, with
$\mathrm{DET}_a\approx0.99$ as high as the ordinary determinism. The chaotic
two-well dynamics therefore still organizes itself around the sign-reversal
symmetry: the antiperiodic fingerprint is not destroyed by the onset of
chaos but becomes statistical rather than exact.

\begin{table*}[t]
\centering
\caption{RQA of the four regimes ($\omega=1.3$, $N=1201$). Left block:
ordinary recurrence plot; right block: anti-recurrence plot (subscript $a$).
$L_\mathrm{max}/N$ separates order from chaos; $\mathrm{RR}_a/\mathrm{RR}$
separates single-well from two-well dynamics.}
\label{tab:rqa}
\begin{ruledtabular}
\begin{tabular}{l c c c c c c c}
Regime & $\gamma$ & $\varepsilon$
 & $\mathrm{RR}$ & $L_\mathrm{max}/N$ & $\mathrm{ENTR}$
 & $\mathrm{RR}_a/\mathrm{RR}$ & $L_{\mathrm{max},a}/N$ \\
\colrule
Periodic, 1 well    & 0.21  & 0.115 & 0.086 & 0.98 & 0.63 & 0.00 & 0.00 \\
Chaotic, 1 well     & 0.36  & 0.156 & 0.053 & 0.34 & 3.22 & 0.00 & 0.00 \\
Antiperiodic, 2 well& 0.637 & 0.284 & 0.051 & 0.93 & 1.93 & 0.94 & 0.04 \\
Chaotic, 2 well     & 0.50  & 0.284 & 0.057 & 0.28 & 2.87 & 0.80 & 0.10 \\
\end{tabular}
\end{ruledtabular}
\end{table*}

\emph{Discussion.}---The pair (RP, anti-RP) classifies the forced
Duffing--Holmes oscillator along two independent axes: through
$L_\mathrm{max}/N$ and $\mathrm{DIV}$ the ordinary RP separates order from
chaos, while the anti-RP---the cross recurrence between the trajectory and
its point reflection, Eq.~\eqref{eq:antiRP}---is sensitive to the dynamical
realization of $\mathcal{S}$ that the ordinary RP cannot see. The ratio
$\mathrm{RR}_a/\mathrm{RR}$ is essentially binary across the four regimes,
near zero for single-well attractors and near unity for the two-well,
symmetry-respecting ones.

The physically significant outcome is the robustness of the antiperiodic fingerprint to the onset of chaos: chaotic attractors that are indistinguishable by standard RQA can nevertheless be differentiated by the anti-RP according to whether they retain the underlying $x\to-x$ symmetry inherited from the antiperiodic orbit. Although exact only for the periodic solution, this symmetry persists in a statistical sense throughout the chaotic dynamics that emerges from it, supporting the view that the antiperiodic family constitutes the organizing skeleton\cite{auerbach1987exploring} of an extended region of parameter space, including its chaotic portions~\cite{marti2026antiperiodic}. Beyond the present case, the anti-RP provides an inexpensive addition to the recurrence toolkit. More generally, it suggests a framework in which recurrence analysis is tailored to specific symmetry operations, providing a systematic way to expose otherwise nontrivial dynamical symmetries in chaotic systems and offering a route toward analogous tools for detecting other classes of symmetries.

\section*{Data availability}
The data that support the findings of this study were generated by
simulation and analysis codes developed by the authors, which build upon
openly available numerical routines distributed through public software
repositories. The codes and the resulting data are available from the
corresponding author upon reasonable request.

\begin{acknowledgments}
A.C.M.\ acknowledges financial support from Project CSIC I+D
\textit{Predictabilidad, caos, regularidad y simetr\'ias en sistemas
f\'isicos no lineales} (No.~22520240100022UD), funded by CSIC--UdelaR.
E.D.L.\ acknowledges Brazilian agencies CNPq (No.~304398/2023-3) and FAPESP
(No.~2021/09519-5, No.~2025/27957-0, No.~2026/00770-0). The authors
acknowledge computing time on the high-performance cluster \texttt{ClusterUY}.
\end{acknowledgments}

\bibliography{rqa}

\end{document}